\documentclass[a4paper,12pt]{article}
\usepackage{amsthm}
\usepackage{amssymb}
\usepackage{amsmath}
\usepackage{amsfonts}

\newtheorem{remark}{Remark}[section]

\newtheorem{lemma}{Lemma}[section]
\newtheorem{theorem}{Theorem}[section]

\newtheorem{corollary}{Corollary}[section]

\def\b1{\mbox{\boldmath $1$}}

\newenvironment{demo*}{\vspace{3mm}\noindent{\bf Proof.}}{\hfill $\Box$ \vspace{3mm}}

\begin{document}

\baselineskip=20pt
\title{\bf \Large
Optimal dividend problem for a generalized  compound Poisson  risk model}
\author{\normalsize{\sc Ying Shen,\ \ Chuancun Yin}   \\
[3mm] {\normalsize\it  School of Mathematical Sciences, Qufu Normal University}\\
{\normalsize\it Shandong 273165, P.R.\ China} \\
 e-mail: ccyin@mail.qfnu.edu.cn} \maketitle
 \centerline{\large {\bf Abstract}}
\vskip0.01cm
In this note we study the  optimal  dividend problem     for a company whose surplus process, in the absence of  dividend payments, evolves as a    generalized  compound Poisson model in which the counting process is a  generalized Poisson process. This model including the classical risk model and  the P\'olya-Aeppli  risk model  as special cases.   The objective is to find a dividend policy so as to maximize the expected discounted value of dividends which are paid to the shareholders until the company is ruined. We show that under some conditions the optimal dividend strategy is formed by a barrier strategy. Moreover, two conjectures are proposed.\\

\noindent {\bf Keywords}: \, {Barrier strategy, Optimal dividend strategy, Generalized  compound Poisson  risk model, Stochastic control.}


\normalsize
\baselineskip=22pt

\section{\normalsize  Introduction}\label{math}
In recent decades, the  optimization problem  of dividend has received remarkable attention in the financial mathematics and insurance mathematics. This  problem goes back to Finetti (1957), who considered a discrete time random walk with step size $\pm 1$ and found that the optimal dividend strategy is a barrier strategy, that is, any surplus above a certain level would be paid as dividend. Since then, many researchers have studied the dividends problems for various risk models under a barrier strategy. For the compound Poisson model, this problem was solved by Gerber (1969), identifying so-called band strategies as the optimal ones. For  exponentially distributed claim sizes this strategy simplifies to a barrier strategy. Azcue and Nuler (2005) follows a viscosity approach to investigate optimal reinsurance and dividend strategies in the Cram\'er-Lundberg model. Albrecher and Thonhauser (2008) showed that the optimality of barrier
strategies in the classical model with exponential claims still holds if there is a constant force of interest.  Avram et al. (2007) considered the case where the risk process is given by a general
spectrally negative L\'evy process and gave a sufficient condition
involving the generator of the L\'evy process for optimality of the
barrier strategy. Loeffen (2008) showed that barrier
strategy is optimal among all admissible strategies for general
spectrally negative L\'evy risk processes with completely monotone
jump density, and Kyprianou et al. (2010)  relaxed this condition on
the jump density to log-convex. An alternative proof is given in
Yin and Wang (2009). Loeffen and Renaud (2010) pushed this result further by assuming the weaker condition that
the L\'evy measure has a density which is log-convex. Azcue and Muler (2010)
examines the analogous questions in the compound Poisson risk model with investment. 
 
 The Poisson processes are the most basic and widely used stochastic model  for modeling discrete data, it may provide a poor fit in the presence of over-dispersion. For example, the use of the Poisson distribution as a model describing the number of claims caused by individual policyholders (e.g. in automobile insurance) during to that certain period is usually rejected, since in practice the
behavior of policyholders is heterogeneous. In such a case the standard Poisson model is inappropriate. For example,
 in collective risk theory, it is assume that claims occur in bulk, where the number of bulks $M_t$ occurring in $(0,t]$ follows a Poisson  process with parameter $\lambda$.  Each bulk consists of a random number of claims so that the total number of claims is of the form
 $N_t=\sum_{i=1}^{M_t}X_i,$ where $\{X_i, i\ge 1\}$ denotes the number of claims in the $i$-th bulk. The aggregate claim payments made up to time $t$,  called the generalized Poisson process, is given by $\sum_{i=1}^{N_t}Y_i,$ where $\{Y_i, i\ge 1\}$ representing the individual claim amounts. In this paper, we formulate and solve an optimal dividends problem for a   generalized Poisson risk model in which the aggregate claim payments is defined by  a generalized Poisson process.
 
The rest of the paper is organized as follows.  In Section 2, we give a rigorous mathematical formulation of the problem. Section 3 gives notion of log-convexity and complete monotonicity. We present our main results in Section 4 and prove them in Section 5.

 \setcounter{equation}{0}
\section{\normalsize  Problem setting}\label{intro}

 Consider the risk model $\{X(t), t\ge 0\}$,
 defined on
the filtered probability space $(\Omega, {\cal{F}},
\Bbb{F}=\{{\cal{F}}_t:t\ge 0\}, P)$ and,
\begin{equation}
X(t)=x+ct+\sigma W_t-\sum_{i=1}^{N_t}Y_i,
\end{equation}
where $\{W_t; t\ge 0\}$ is a standard Brownian motion with $W_0=0$, the claim sizes
$\{Y_i; i\ge 1\}$ are positive independent and identically distributed random
variables whose probability distribution function is given by $P(y)$,
$\{N_t; t\ge 0\}$ is a generalized Poisson process defined as 
$N_t=\sum_{i=1}^{M_t}X_i,$ where $\{X_i, i\ge 1\}$ are discrete independent and identically distributed random
variables whose probability  distribution is given by $P(X_i=k)=p_k, k=1,2,\cdots,$ and $\{M_t, \ge 0\}$ is a  homogeneous Poisson process with intensity $\lambda>0$.
Moreover,   it is assumed that $\{W_t\}, \{M_t\}$,  $\{X_i\}$   and $\{Y_i\}$ are mutually
independent.
In particular, when $P(X_i=1)=1$, the process $\{N_t\}$ reduces to the homogeneous
 Poisson process with intensity $\lambda>0$, and hence the risk model (2.1) reduces to the classical risk model perturbed by Brownian motion (see Chiu and Yin (2003)).
 
 The probability mass function of $N_t$ is given by
 \begin{equation}
 P(N_t=n)=\sum_{k=0}^{\infty}e^{-\lambda t}\frac{(\lambda t)^k}{k!}p_k^{*n},\ n=0, 1,2 \cdots,
 \end{equation}
 where $p_{k}^{*n}$ is the $n$-fold convolution of $\{p_k\}$.
 In a few special cases it is possible to determine the probabilities $P(N_t=n)$'s explicitly.

 {\bf Example 2.1} \ Suppose that $X_1, X_2,\cdots$  are geometrically distributed with parameter $1-\rho$, where
 $\rho\in (0,1)$, i.e.
 $$P(X_i=k)=(1-\rho){\rho}^{k-1}, k=1,2,\cdots.$$
 Then the compound Poisson process by geometric compounding leads to   the P\'olya-Aeppli process $\{N_t, t\ge 0\}$  with
parameters $\lambda$ and $\rho$ (cf. Minkova (2004)). That is for all $t\ge 0$,
\begin{equation}
  P(N_t=n)=\left\{
  \begin{array}{ll}e^{-\lambda t}, & {\rm if} \ n=0,\\
  e^{-\lambda t}\sum\limits_{i=1}^{n}\left(\begin{array}{c}n-1\\i-1\end{array}\right)\frac{[\lambda(1-\rho)t]^i}{i!}{\rho}^{n-i},
   &{\rm if} \ n=1,2,\cdots.
  \end{array}
  \right.
\end{equation}
Note that the P\'olya-Aeppli process is a time-homogeneous process, it is also called Poisson-geometric process in Chinese literature, for example see Mao and Liu (2005), where the ruin probability was studied for compound Poisson-geometric process. In the case of $\rho=0$, the P\'olya-Aeppli process becomes a homogeneous Poisson process.

{\bf Example 2.2} (Quenouille (1949))\  Let $\{X_i, i\ge 1\}$ denote a sequence of independent and identically distributed random variables, each one having the logarithmic distribution (also known as the logarithmic series distribution) $\ln(\theta)$, with probability mass function
$$P(X_i=n)=\frac{\theta^{n}}{-n\ln (1-\theta)}, \ n=1,2, \cdots, 0<\theta<1.$$
 Suppose that $M_t$ has a Poisson process with parameter $\lambda=-r\ln(1-\theta)$. Then the random sum
$$N_t=\sum_{i=1}^{M_t}X_i,$$
has the negative binomial distribution $NB(rt,\theta)$:
 \begin{equation}
 P(N_t=n)=\left(\begin{array}{c}n+rt-1\\n\end{array}\right)(1-\theta)^{rt}\theta^n,\  n=0,1,2, \cdots.
 \end{equation}
 In this way, the negative binomial distribution is seen to be a compound Poisson distribution.

We now consider the classical optimal dividend control problem.
Let $\pi$ be a dividend strategy consisting of a non-decreasing left-continuous $\Bbb{F}$-adapted process $\pi=\{L_t^{\pi}, t\ge 0\}$ with $L_0^{\pi}=0$, where $L_t^{\pi}$ represents the cumulative dividends paid out by the company till time $t$ under the control $\pi$. We define the controlled risk process $U^{\pi}=\{U_t^{\pi}, t\ge 0\}$ by $U_t^{\pi}=X(t)-L_t^{\pi}$. Let $\tau^{\pi}=\inf\{t>0: U_t^{\pi}<0\}$ be the ruin time and define the value function of a dividend strategy $\pi$ by
$$V_{\pi}(x)=E\left[\int_0^{\tau_{\pi}}e^{-q t}dL^{\pi}(s)|U_0^{\pi}=x\right],$$
 where $q>0$  is an interest force for the calculation of the present value. Let $\Xi$ be the set of all admissible dividend  strategies, that is all strategies $\pi$ such that
 $L_{t+}^{\pi}-L_{t}^{\pi}\le U_t^{\pi}$ for $t<\tau^{\pi}$.
The objective is to   solve  the following stochastic control problem:
\begin{equation}
V(x)=\sup_{\pi\in\Xi}V_{\pi}(x),
\end{equation}
and to find an optimal policy $\pi^*\in\Xi$ that satisfies $V(x)=V_{\pi^*}(x)$  for all $x\ge 0$.

 \setcounter{equation}{0}
\section{\normalsize   Log-convexity and complete monotonicity}\label{intro}

Before starting our main results, we introduce  the definitions of log-convexity and complete monotonicity.

{\bf Definition 3.1.} (Willmot and Lin (2001)). (1) \  A distribution $\{P_n\}$ on the non-negative integers is said to be log-convex if
$P_n^2\le P_{n+1}P_{n-1}, n=1,2,\cdots,$ and $\{P_n\}$ is said to be strictly log-convex if
$P_n^2<P_{n+1}P_{n-1}, n=1,2,\cdots.$
 A counting distribution $\{r_n, n\ge 0\}$ is discrete  completely
monotone iff it is a mixture of geometric distributions, i.e.
$$r_n=\int_0^1 (1-\theta)\theta^n dU(\theta),$$
where $U$ is a probability distribution on $(0,1)$.\\
(2)\ A function $f: \Bbb{R}\rightarrow \Bbb{R}^+$ is
log-convex if $\log f(x)$ is a convex function. Let  $f\in C^{\infty}(0,\infty)$ with
$f\ge 0$. We say $f$  is completely monotone if  $(-1)^n f^{(n)}\ge
0$ for all $n\in \Bbb{N}$.

(3)\ The distribution function $G(x)$ is said to be decreasing (increasing) failure rate or DFR (IFR) if $\overline{G}(x+y)/\overline{G}(y)$ is nondecreasing (nonincreasing) in $y$ for fixed $x\ge 0$, i.e. if
$\overline{G}(y)$ is log-convex (log-concave).

Note that the completely monotone class is a subclass of the log-convex. For examples of continuous log-convex or completely monotone functions can be found in Yin and Wang (2009). Now, we give a discrete example.

{\bf Example 3.1}\  Let $N$ be a logarithmic random variable with
$$p_n=P(N=n)=\frac{\theta^{n+1}}{-(n+1)\log (1-\theta)}, \ n=0,1,2, \cdots, 0<\theta<1.$$
Then $\{p_n\}$ is completely monotone (see van Harn (1978, P. 58)).
The  generalized logarithmic  series distribution is defined by
$$r_n=\frac{1}{\beta n}\frac{\Gamma(\beta n+1)}{\Gamma (\beta n -n+1)\Gamma (n+1)}\theta^n(1-\theta)^{\beta n-n}/(-\log (1-\theta)),\ n=1,2,\cdots,$$
with $\beta\ge 1$ and $0<\theta<\beta^{-1}$.  Then $\{r_n, n\ge 1\}$   is strictly log-convex (see Hansen and Willekens (1990)).

\section{\normalsize  Main results}\label{intro}

Denote by $\pi_b=\{L_t^b, t\ge 0\}$ the constant barrier strategy at level $b$ which is defined by $L_0^b=0$ and
$$L_t^b=\left(\sup_{0\le s<t}X(s)-b\right)\vee 0$$ for all $t>0$.
That is, for a level $b>0$   whenever surplus goes above
$b$, the excess is  paid as dividends to the shareholders of the
company and, if the surplus is
less than $b$, no dividends are paid out. We will now present the main results of this note which give sufficient conditions for optimality of a barrier strategy $\pi_{b^*}$. It is  important to note that various dividend strategies can be employed by an
insurance company. However, we will only focus on the conditions for the optimality of a dividend  strategy.

After some tedious calculations, we get
$$\psi(s):=\ln Ee^{s X(1)}=cs+\frac12 \sigma^2 s^2+\lambda\int_0^{\infty}(e^{-sz}-1)dF(z), \Re(s)\ge 0,$$
where
\begin{equation}
F(z)=\sum_{k=1}^{\infty}p_k P^{*k}(z).
 \end{equation}
 Here   $P^{*k}$ is the $k$-fold convolution of $P$ with itself.
So that $X$ is a special spectrally negative L\'evy process with the Laplace exponent $\psi(s)$.
  Therefore, all the known results for   spectrally negative L\'evy process models can be applied to the model (2.1). However, since the distribution function $F$  is not  explicit (depends on the distribution of $Y_i$ and $X_i$), it can be of interest to study which assumptions on the probability distributions of $Y_i$ and $N_t$ ensure that the optimal dividend strategy is barrier one.
  
  We now recall the definition of the $q-$scale function $W^{(q)}$ and some properties of this function. For  each $q\ge 0$ there exists a
continuous and increasing function $W^{(q)}:\Bbb{R}\rightarrow
[0,\infty)$, called the $q$-scale function defined in such a way
that $W^{(q)}(x) = 0$ for all $x < 0$ and on $[0,\infty)$ its
Laplace transform is given by
\begin{equation}
\int_0^{\infty}\text{e}^{-s
x}W^{(q)}(x)dx=\frac{1}{\psi(s)-q},\; s
>\rho(q).\nonumber
\end{equation}
Here, $\rho(q)$ is the unique root of equation $\psi(s)-q=0$ in the half-plane $\Re(s)\ge 0$.

From Avram et al. (2007 ) we get the expected discounted
value of dividend payments of the barrier strategy at level $b\ge 0$ is given by
\begin{equation}
 V_b(x)=\left\{
  \begin{array}{ll} \frac{W^{(q)}(x)}{{W^{(q)}}'(b)}, & {\rm if} \ 0\le x\le b,\\
    x-b+ \frac{W^{(q)}(b)}{{W^{(q)}}'(b)}, & {\rm if} \ x>b.
  \end{array}
  \right.
\end{equation}

Define
$$b^*=\{b\ge 0: {W^{(q)}}'(b)\le {W^{(q)}}'(x), x\ge 0\}.$$
 
\begin{theorem} For model (2.1),  if  $P$ has a completely monotone probability density function on  $(0,\infty)$
 and $\{p_n, n\ge 0\}$ is discrete  completely monotone,  then   the
 barrier strategy with  level $b^*$ is the optimal dividend strategy. 
 Moreover, the $V$ defined by (2.5) is given by $V(x)=V_{b^*}(x)$.
 \end{theorem}
\begin{theorem} For model (2.1), if $\{p_n, n\ge 1\}$ is discrete  completely monotone and   $P$ is  DFR,   then   the
 barrier strategy with  level $b^*$ is the optimal dividend strategy.
 Moreover, the $V$ defined by (2.5) is given by $V(x)=V_{b^*}(x)$.
\end{theorem}
\begin{corollary} For model (2.1) with $N_t$ given by (2.3) or (2.4), if   $P$ is  DFR,   then   the
 barrier strategy with  level $b^*$ is the optimal dividend strategy.
 Moreover, the $V$ defined by (2.5) is given by $V(x)=V_{b^*}(x)$.
\end{corollary}
\begin{theorem} For model (2.1),  if $\{p_n, n\ge 1\}$ is  a log-convex probability mass function  and $P$  is the exponential distribution function with mean $1/\beta$,   then   the
 barrier strategy with  level $b^*$ is the optimal dividend strategy.
 Moreover, the $V$ defined by (2.5) is given by $V(x)=V_{b^*}(x)$.
\end{theorem}

\setcounter{equation}{0}
\section{Proof of main results}\label{Thr}

Before proving the main results, we give several lemmas.
\begin {lemma}{(Loeffen (2008)} Suppose that the L\'evy measure  of a spectrally negative L\'evy process $X$ has a completely monotone density on $(0,\infty)$,  then  the barrier strategy at $b^*$ is an optimal strategy.
\end {lemma}

Kyprianou,  Rivero and Song (2010) providing weaker conditions on the L\'evy
measure for the optimality of a barrier strategy. An alternative approach can be found in Yin and Wang (2009).
\begin{lemma}   Suppose that a spectrally negative L\'evy process $X$ has a
L\'evy  density $\pi$ on  $(0,\infty)$ that is log-convex, then  the barrier strategy at $b^*$ is an optimal
strategy.
\end{lemma}
Note that for  the Cram\'er-Lundberg model with or without a Brownian component, the requirement of log-convexity of
the L\'evy  density $\pi$ on  $(0,\infty)$ is equivalent to the log-convexity of the  probability density function of the individual claim amount  on $(0,\infty)$.
Since the L\'evy measure having a log-convex (or completely monotone) density implies that tail of the  L\'evy measure is log-convex and the converse is not true (cf. Loeffen and Renaud (2010)), the following result improves the results in Lemmas 3.1 and 3.2.
\begin{lemma} (Loeffen and Renaud (2010))   Suppose that the  tail of the  L\'evy measure  of a spectrally negative L\'evy process $X$ is log-convex, then  the barrier strategy at $b^*$ is an optimal strategy.
\end{lemma}

{\bf Proof of Theorem 4.1.}\  If $\{p_n, n\ge 1\}$ is discrete  completely monotone and $P$ has a completely monotone density on $(0,\infty)$, then
\begin{equation}
F(z)=\sum_{k=1}^{\infty}p_k P^{*k}(z)\nonumber
 \end{equation}
 has a completely monotone density on $(0,\infty)$ (cf. Chiu and Yin (2013)). The result follows from Lemma 5.1.

{\bf Proof of Theorem 4.2.}\; It is well known that the property of DFR is preserved under the geometric sum (see Shanthikumar (1988, Corollary 3.6)), and since the sum of two log-convex functions   is  log-convex  and the limit of a pointwise convergent sequence of log-convex functions   is  log-convex, it follows that
\begin{equation}
F(z)=\sum_{k=1}^{\infty}p_k P^{*k}(z)\nonumber
 \end{equation}
 is also DFR.
The result of   Theorem 4.2 follows from Lemma 5.3.

{\bf Proof of Theorem 4.3.}\; If $P$  is the exponential distribution function with mean $1/\beta$, then by (4.1) we have
$$F(z)=\sum_{k=1}^{\infty}p_k \left(1-e^{-\beta z}\sum_{j=0}^{k-1}\frac{(\beta z)^j}{j!}\right).$$
Therefore,
$$\overline{F}(z)=\sum_{k=1}^{\infty}p_k \left(e^{-\beta z}\sum_{j=0}^{k-1}\frac{(\beta z)^j}{j!}\right).$$
Interchanging the order of summation yields
$$\overline{F}(z)=e^{-\beta z}\sum_{j=0}^{\infty}\overline{P}_j \frac{(\beta z)^j}{j!},$$
where
$$\overline{P}_j=\sum_{i=j+1}^{\infty}p_i.$$
Note that $1=\overline{P}_0 \ge \overline{P}_1\ge \overline{P}_2\ge \cdots$ and $p_{k+1}/p_k$ is increasing in $k$,
it follows from Theorem 3.2 in Esary and Marshall (1973) that $F$  has a density which is logarithmically convex
on $(0,\infty)$. The result  follows from Lemmas 5.2.
\begin{remark}
\ At the end of this paper, we give two conjectures. The first  conjecture can be viewed as an extension of
Theorem 4.3; The second  conjecture can be viewed as an extension of Conjecture 1 and
Theorem 4.2.

{\bf Conjecture 1}.\;  For model (2.1),  if $\{p_n, n\ge 1\}$ is  a log-convex   and $P$
has a density $\pi$ on  $(0,\infty)$ that is log-convex, then  the barrier strategy at $b^*$ is an optimal
strategy for stochastic control problem (2.5).

{\bf Conjecture 2}.\;  For model (2.1),  if $\{p_n, n\ge 1\}$ is DFR and   $P$ is  DFR, then  the barrier strategy at $b^*$ is an optimal strategy for stochastic control problem (2.5).
\end{remark}

\vskip0.3cm

\noindent{\bf Acknowledgements}\;
The research was supported by the National Natural
Science Foundation of China (No.11171179), the Research Fund for
the Doctoral Program of Higher Education of China (No. 20133705110002) and the Program for  Scientific Research Innovation Team
in Colleges and Universities of Shandong Province.

\end{document}